\documentclass[manuscript]{aastex631}
\usepackage[utf8]{inputenc}
\usepackage{amsmath}
\usepackage{lipsum}
\usepackage{color,soul}
\usepackage{appendix}
\usepackage{tikz}

\begin{document}

\title{Detection of Travel Time Anisotropy from Subsurface Horizontal Magnetic Fields}

\correspondingauthor{John T. Stefan}
\email{jts25@njit.edu}

\author[0000-0002-5519-8291]{John T. Stefan}
\affiliation{Department of Physics, New Jersey Institute of Technology, Newark, NJ 07102}

\author[0000-0003-0364-4883]{Alexander G. Kosovichev}
\affiliation{Department of Physics, New Jersey Institute of Technology, Newark, NJ 07102}
\affiliation{NASA Ames Research Center, Moffett Field, Mountain View, CA 94040}

\begin{abstract}
    A time-distance measurement technique is derived to isolate phase travel time anisotropy caused by subsurface horizontal magnetic fields, and a method which uses the measured anisotropy to estimate the field's orientation is also derived. A simulation of acoustic waves propagating in a uniform, inclined magnetic field with solar background structure is used to verify the derived technique. Then, the procedure is applied to a numerical simulation of a sunspot, for which the subsurface state is known, to provide context for the results obtained from the study of several sunspots observed by the Helioseismic and Magnetic Imager. Significant anisotropies are detected, on the order of one minute, and the subsurface field's azimuth is estimated and compared with the azimuth of the surface magnetic field. In all cases, the subsurface azimuth is found to be well-aligned with that of the surface, and the results from the numerical simulation are used to interpret features in the detected travel time anisotropy. 
\end{abstract}

\section{Introduction}

Improvements in local helioseismology over the past few decades have made measurements of the solar interior more accessible and reproducible, with regards to travel time deviations produced by sound speed perturbations and horizontal flows in particular, and recent work now focuses on developing better inversion methods \citep{cs_verification,Hanasoge_inv,Jackiewicz}. The measurements of flows are useful for the study of the Sun's meridional circulation and differential rotation, especially in their relation to the solar dynamo \citep{getling_circ,dynamo_connec}, and applications of detected sound speed perturbations include the structure of sunspots \citep{tomog} and identification of emerging magnetic flux \citep{Zharkov, Ilonidis}. Though these applications are certainly worthwhile, there has yet to be any recent developments in the measurement of new quantities.

As one of the more dramatic manifestations of the solar dynamo, active regions (ARs) remain a popular object of study. Helioseismic investigations of these structures is generally limited to their surrounding flows \citep{Jain_spotflows,Braun_flows} and subsurface sound speed variations \citep{Zhao_cs}. In this work, we introduce a new measurement technique to detect subsurface horizontal magnetic fields, a novel quantity in the helioseismology of active regions. The basis of this detection is the conversion of purely acoustic waves into fast magnetoacoustic waves in the presence of strong magnetic fields. \cite{Cally} find significant conversion at the equipartition surface - where the Alfv\'en and sound speeds are equal - for vertical magnetic fields, and \cite{Jain_bh} find that horizontal magnetic fields produce a phase speed dependent perturbation to acoustic travel times in the ray path approximation. \cite{Jain_bh} further states that disentangling travel time perturbations produced by magnetic fields from those produced by sound speed and other structural variations may be difficult. In Section \ref{deriv}, we show that the measurement scheme reduces contributions from non-magnetic effects.

Aside from the difficulty in isolating the magnetic contribution to the acoustic travel time, one must also contend with the reduction of acoustic power in active regions, particularly in waves propagating from the active region outwards \citep{ac_power}. We attempt to overcome this by extending the observation window from 8 hours, as is used in most helioseismic investigations, to 24 hours. This extended window restricts the active regions that can be studied to those where the magnetic field and morphology remain constant throughout the observations. We first verify the derived equations by applying the derived measurement technique to the simple case of an acoustic point source in a uniform, inclined magnetic field. Then, a numerical model of a sunspot, for which most of the subsurface characteristics are known and are approximately constant, is investigated. Finally, we apply the time-distance procedure to three active regions observed by the Helioseismic and Magnetic Imager (HMI) instrument aboard the Solar Dynamics Observatory (SDO). 

\section{Methods and Data}

The general procedure for measuring the travel time anisotropy begins by tracking a given area of interest and remapping the Dopplergram series to a Postel's projection with a spatial resolution of 0.12 degrees per pixel. Each image has size 256 by 256 pixels, and the series spans 24 hours at a cadence of 45 seconds. The data is then treated with a phase speed filter whose parameters are the same as those used in the Helioseismic and Magnetic Imager (HMI) time-distance pipeline \citep{TDpipe}. The purpose of the phase speed filter is to isolate acoustic waves whose turning points fall within a desired range of depth. Then for each pixel in the image, we cross-correlate the Doppler signal of the pixel with the spatial average of several arcs placed around the center. The cross-correlations are then processed to obtain two travel times, the first from Gabor wavelet fitting \citep{Nigam} and the second from the travel time definition of \cite{GB02}. The spatial averaging scheme, as well as how the acoustic travel times can be used to obtain information about the subsurface horizontal magnetic field, are detailed in the following sections.

\subsection{Obtaining the Angle and Magnitude of the Horizontal Magnetic Field from Acoustic Travel Times}\label{deriv}
In general, a perturbation to an oscillation mode's wavenumber, $\delta k$, will produce a corresponding perturbation to the mode's travel time, $\delta \tau$, given by

\begin{equation*}\label{dt}
\delta\tau=\int_{\Gamma}\dfrac{\delta k}{\omega}ds,
\end{equation*}
where $\Gamma$ is the unperturbed ray path, $\omega$ is the unperturbed frequency of the mode, and $s$ is the arc length along the unperturbed ray path. Given wavenumber perturbations produced by perturbations to sound speed, to the acoustic cut-off frequency, and from the presence of subsurface flows and magnetic fields, the perturbed one-way travel time is
\begin{equation}\label{dtfull}
\delta\tau=-\int_{\Gamma}\left[\dfrac{\mathbf{n}\cdot\mathbf{U}}{c^{2}}+\dfrac{\delta c}{c}S+\dfrac{\delta \omega_c}{\omega_c}\dfrac{\omega_c^2}{\omega^2c^2S}+\dfrac{1}{2}\left(\dfrac{|\mathbf{c_{A}}|^{2}}{c^2}-\dfrac{\left(\mathbf{k}\cdot\mathbf{c_{A}}\right)^2}{|\mathbf{k}|^{2}c^2}\right)S\right]ds,
\end{equation}
where $\omega_c$ is the acoustic cut-off frequency, $\mathbf{U}$ represents the background flow, $c$ is the sound speed, $\mathbf{c_{A}}=\mathbf{B}/\sqrt{4\pi\rho}$ is the Alfv\'en speed, $k$ is the complete wavevector, $\mathbf{n}=\mathbf{k}/|\mathbf{k}|$ is the direction of propagation, and $S=k/\omega$ is the phase slowness. A more complete description of the non-magnetic terms is given by \cite{tomog}, and the derivation of the wavenumber perturbation produced by the presence of a magnetic field is described in Appendix \ref{mhd_deriv}.

Depending on the source of the perturbation, acoustic waves may have a different travel time in one direction along the ray path than waves traveling in the opposite direction. This can be used to isolate particular quantities in equation \ref{dtfull} for a more accurate estimation of their magnitude. Taking the difference of the travel times of acoustic waves traveling in opposite directions along the same ray path isolates the flow term, while taking the mean of the opposing travel times isolates the sound speed, cut-off frequency, and magnetic terms; we express the latter as
\begin{equation*}\label{dtmag_simp}
\delta\tau_{\text{mean}}=T_{A}-\dfrac{1}{2}\int_{\Gamma}\left(\dfrac{|\mathbf{c_{A}}|^{2}}{c^2}-\dfrac{\left(\mathbf{k}\cdot\mathbf{c_{A}}\right)^2}{|\mathbf{k}|^{2}c^2}\right)Sds,
\end{equation*}
where the acoustic terms are represented by $T_{A}$.

We consider the magnetic field to have a radial component $B_r$ and a horizontal component $B_{h}=\sqrt{B_\theta^{2}+B_{\phi}^2}$. We label the mean travel time of an acoustic ray traveling along the North-South axis as $\delta\tau_{\text{mean}}^{(1)}$, given by
\[
\delta\tau_{\text{mean}}^{(1)}=T_{A}-\dfrac{1}{2}\int_{\Gamma}\dfrac{1}{|\mathbf{k}|^{2}c^{2}}\left[\dfrac{1}{4\pi\rho}\left(k_{r}^{2}+k_{h}^{2}\right)\left(B_{r}^{2}+B_{h}^{2}\right)-\dfrac{1}{4\pi\rho}\left(k_{r}B_{r}+k_{h}B_{h}\cos(\alpha)\right)^{2}\right]Sds.
\]
Here, the ray is traveling entirely in the $\hat{\theta}$ direction, so only the radial and polar-angle terms remain in the dot product, and $\alpha$ is the angle between the horizontal magnetic field and the North-South axis - the magnetic field's azimuth. Assuming the magnetic field is uniform along the ray path, the expected perturbation to the travel time of a North-South traveling acoustic wave is then
\begin{align}\label{antisymm}
\begin{split}
\delta\tau_{\text{mean}}^{(1)}=T_{A}-\dfrac{1}{2}\int_{\Gamma}\dfrac{1}{|\mathbf{k}|^{2}c^{2}} & \left(k_{r}^{2}c_{A,h}^{2}+k_{h}^{2}c_{A,r}^{2}\right)Sds-\left(\dfrac{1}{2}\int_{\Gamma}\dfrac{1}{|\mathbf{k}|^{2}c^{2}}k_{h}^{2}c_{A,h}^{2}Sds\right)\sin^{2}(\alpha) \\ & +\left(\int_{\Gamma}\dfrac{1}{|\mathbf{k}|^{2}c^{2}}k_{r}k_{h}c_{A,r}c_{A,h}Sds\right)\cos(\alpha).
\end{split}
\end{align}
This can be further reduced by noting that the last term is antisymmetric along the ray path, which originates from the sign flip of $k_r$ at the mode's turning point. We then have
\begin{equation}\label{NS_dt}
\delta\tau_{\text{mean}}^{(1)}=T_{A}-\dfrac{1}{2}\int_{\Gamma}\dfrac{1}{|\mathbf{k}|^{2}c^{2}}\left(k_{r}^{2}c_{A,h}^{2}+k_{h}^{2}c_{A,r}^{2}\right)Sds-\left(\dfrac{1}{2}\int_{\Gamma}\dfrac{1}{|\mathbf{k}|^{2}c^{2}}k_{h}^{2}c_{A,h}^{2}Sds\right)\sin^{2}(\alpha).
\end{equation}

We further consider acoustic rays traveling along the Northeast-Southwest direction ($\delta\tau_{\text{mean}}^{(2)}$), along the East-West direction ($\delta\tau_{\text{mean}}^{(3)}$), and along the Southeast-Northwest direction ($\delta\tau_{\text{mean}}^{(4)}$); each of these is oriented an additional $\pi/4$ radians from the previous ray. In each case, we re-orient our coordinate system so that the ray is always traveling in the local $\hat{\theta}$ direction, and the orientation of the horizontal magnetic field receives a corresponding adjustment of $\pi/4$ radians. Performing a similar reduction as for the North-South traveling ray, the expected travel time perturbations to these rays are
\begin{equation}\label{NESW_dt}
\delta\tau_{\text{mean}}^{(2)}=T_{A}-\dfrac{1}{2}\int_{\Gamma}\dfrac{1}{|\mathbf{k}|^{2}c^{2}}\left(k_{r}^{2}c_{A,h}^{2}+k_{h}^{2}c_{A,r}^{2}\right)Sds-\left(\dfrac{1}{2}\int_{\Gamma}\dfrac{1}{|\mathbf{k}|^{2}c^{2}}k_{h}^{2}c_{A,h}^{2}Sds\right)\sin^{2}\left(\alpha+\dfrac{\pi}{4}\right),
\end{equation}
\begin{equation}\label{EW_dt}
\delta\tau_{\text{mean}}^{(3)}=T_{A}-\dfrac{1}{2}\int_{\Gamma}\dfrac{1}{|\mathbf{k}|^{2}c^{2}}\left(k_{r}^{2}c_{A,h}^{2}+k_{h}^{2}c_{A,r}^{2}\right)Sds-\left(\dfrac{1}{2}\int_{\Gamma}\dfrac{1}{|\mathbf{k}|^{2}c^{2}}k_{h}^{2}c_{A,h}^{2}Sds\right)\cos^{2}(\alpha),
\end{equation}
and
\begin{equation}\label{SENW_dt}
\delta\tau_{\text{mean}}^{(4)}=T_{A}-\dfrac{1}{2}\int_{\Gamma}\dfrac{1}{|\mathbf{k}|^{2}c^{2}}\left(k_{r}^{2}c_{A,h}^{2}+k_{h}^{2}c_{A,r}^{2}\right)Sds-\left(\dfrac{1}{2}\int_{\Gamma}\dfrac{1}{|\mathbf{k}|^{2}c^{2}}k_{h}^{2}c_{A,h}^{2}Sds\right)\cos^{2}\left(\alpha+\dfrac{\pi}{4}\right).
\end{equation}

Subtracting equation \ref{EW_dt} from equation \ref{NS_dt}, and equation \ref{SENW_dt} from equation \ref{NESW_dt}, yields
\begin{align}\label{diff1}
\begin{split}
\delta\tau_{\text{mean}}^{(1)}-\delta\tau_{\text{mean}}^{(3)} & =-\left(\dfrac{1}{2}\int_{\Gamma}\dfrac{1}{|\mathbf{k}|^{2}c^{2}}k_{h}^{2}c_{A,h}^{2}Sds\right)\left(\sin^2(\alpha)-\cos^{2}(\alpha)\right) \\
& =\left(\dfrac{1}{2}\int_{\Gamma}\dfrac{1}{|\mathbf{k}|^{2}c^{2}}k_{h}^{2}c_{A,h}^{2}Sds\right)\cos(2\alpha)
\end{split}
\end{align}
and
\begin{align}\label{diff2}
\begin{split}
\delta\tau_{\text{mean}}^{(2)}-\delta\tau_{\text{mean}}^{(4)} & =-\left(\dfrac{1}{2}\int_{\Gamma}\dfrac{1}{|\mathbf{k}|^{2}c^{2}}k_{h}^{2}c_{A,h}^{2}Sds\right)\left(\sin^2\left(\alpha+\dfrac{\pi}{4}\right)-\cos^{2}\left(\alpha+\dfrac{\pi}{4}\right)\right) \\
& =-\left(\dfrac{1}{2}\int_{\Gamma}\dfrac{1}{|\mathbf{k}|^{2}c^{2}}k_{h}^{2}c_{A,h}^{2}Sds\right)\sin(2\alpha).
\end{split}
\end{align}

Finally, combining equations \ref{diff1} and \ref{diff2} in the following ways will yield expressions for the orientation of the horizontal magnetic field
\begin{equation}\label{angle}
\alpha=\dfrac{1}{2}\tan^{-1}\left(-\dfrac{\delta\tau_{\text{mean}}^{(2)}-\delta\tau_{\text{mean}}^{(4)}}{\delta\tau_{\text{mean}}^{(1)}-\delta\tau_{\text{mean}}^{(3)}}\right),
\end{equation}
as well as for the magnetic anisotropy parameter
\begin{equation}\label{aniso}
A \equiv \int_{\Gamma}\dfrac{1}{|\mathbf{k}|^{2}c^{2}}k_{h}^{2}c_{A,h}^{2}Sds=2\sqrt{\left(\delta\tau_{\text{mean}}^{(2)}-\delta\tau_{\text{mean}}^{(4)}\right)^2+\left(\delta\tau_{\text{mean}}^{(2)}-\delta\tau_{\text{mean}}^{(4)}\right)^2},
\end{equation}
which is a direct measure of the magnitude of the horizontal magnetic field. It is important to note that analytically, the position of the minus sign in equation \ref{angle} is irrelevant, but must be attributed to the numerator in numerical computations.

\subsection{Dopplergram Phase Speed Filtering}

We first begin by applying a Gaussian phase speed filter to the tracked Dopplergram series for a given area. As previously mentioned, the parameters for the filter are the same as those used in the HMI time-distance pipeline. As an example, to isolate waves whose turning points $z_t$ lie between 3 and 5 Mm beneath the photosphere, a central phase speed of $v_{\text{phase}}=4.9$ $\mu\text{Hz }l^{-1}$ is used, with a full-width half-maximum (FWHM) of $1.25$ $\mu\text{Hz }l^{-1}$, where $l=k_h R_{\odot}$ is the angular degree of a given mode's oscillation. We refer to modes with the desired turning point depths as belonging to a given annulus group, as the horizontal distance traveled by these modes fall within a particular range. For the previously described phase speed filter, these modes are part of annulus group 3 with horizontal travel distances between 1.08 and 1.32 heliographic degrees.

In our work, we will consider annulus groups 4 ($z_t=5-7$ Mm), 5 ($z_t=7-10$ Mm), and 6 ($z_t=10-13$ Mm), and the relevant parameters are detailed in Table \ref{params}. Annulus groups 1 ($z_t=0-1$ Mm), 2 ($z_t=1-3$ Mm), and 3 generally have travel times too noisy to reliably probe the subsurface magnetic field. The acoustic waves in annulus groups 7 ($z_t=13-17$ Mm) to 11 ($z_t=30-35$ Mm) are typically too deep for the perturbations in travel time caused by the subsurface magnetic field to be resolved, as the effects of acoustic perturbations become stronger than those from the magnetic field with increasing depth.

\begin{table}[h]
    \begin{center}
    \resizebox{\textwidth}{!}{
    \begin{tabular}{||c|c|c|c|c||}
         \hline 
         Annulus Group & Depth Range (Mm) & Horizontal Travel Distance (deg) & Phase Velocity (km s$^{-1}$)  & Gaussian Width (km s$^{-1}$)\\
         \hline 
         \hline 
         4 & 5-7 & 1.44-1.80 & 28.83 & 9.40\\
         \hline
         5 & 7-10 & 1.92-2.40 & 36.48 & 5.91\\
         \hline
         6 & 10-13 & 2.40-2.88 & 40.25 & 5.17\\
         \hline
    \end{tabular}}
    \caption{Parameters for annulus groups 4, 5, and 6. The applied Gaussian phase speed filter is centered at the listed phase speed and width.}\label{params}
    \end{center}
\end{table}

\subsection{Measuring the Acoustic Travel Times}

\begin{figure}[ht!]
\begin{center}
\begin{tikzpicture}

\draw [ultra thick,blue,domain=-22.5:22.5] plot ({2*cos(\x)}, {2*sin(\x)});
\draw [ultra thin,black,rotate=0] (0,0) -- node[above,fill opacity=1] { } (2cm,0);

\draw [ultra thick,red,domain=22.5:67.5] plot ({2*cos(\x)}, {2*sin(\x)});
\draw [ultra thin,black,rotate=45] (0,0) -- node[right,fill opacity=1] { } (2cm,0);

\draw [ultra thick,green,domain=67.5:112.5] plot ({2*cos(\x)}, {2*sin(\x)});
\draw [ultra thin,black,rotate=90] (0,0) -- node[right,fill opacity=1] { } (2cm,0);

\draw [ultra thick,orange,domain=112.5:157.5] plot ({2*cos(\x)}, {2*sin(\x)});
\draw [ultra thin,black,rotate=135] (0,0) -- (2cm,0);

\draw [ultra thick,blue,domain=157.5:202.5] plot ({2*cos(\x)}, {2*sin(\x)});
\draw [ultra thin,black,rotate=180] (0,0) -- (2cm,0);

\draw [ultra thick,red,domain=202.5:247.5] plot ({2*cos(\x)}, {2*sin(\x)});
\draw [ultra thin,black,rotate=225] (0,0) -- (2cm,0);

\draw [ultra thick,green,domain=247.5:292.5] plot ({2*cos(\x)}, {2*sin(\x)});
\draw [ultra thin,black,rotate=270] (0,0) -- (2cm,0);

\draw [ultra thick,orange,domain=292.5:337.5] plot ({2*cos(\x)}, {2*sin(\x)});
\draw [ultra thin,black,rotate=315] (0,0) -- node[right,fill opacity=1] { } (2cm,0);

\filldraw[fill=black,draw=black] (0,0) circle[radius=0.05cm];
\end{tikzpicture}
\caption{Spatial averaging scheme used to measure associated travel times. The Doppler signal within an arc is averaged and cross-correlated with the center pixel, for the mean North-South travel time (green), the mean Northeast-Southwest travel time (orange), the mean East-West travel time (blue), and the mean Southeast-Northwest travel time (red).}\label{measure}
\end{center}
\end{figure}
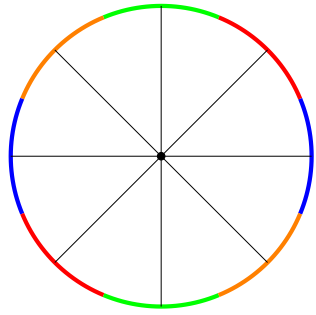

 We divide each annulus group into three separate ranges, one for the minimum horizontal travel distance of the group, one for the mean horizontal travel distance of the group, and one for the maximum horizontal travel distance of the group. Once the Dopplergram series has been filtered, we perform a series of cross-correlations for each pixel with arcs placed at a distance corresponding to a given annulus group's minimum horizontal travel distance. The arcs each cover 45 degrees and are placed according to the desired travel time measurements (i.e. North-South, East-West, etc.), and a schematic is provided in Figure \ref{measure}. Doppler signal within a half-pixel's distance from these arcs is averaged so that the arc can be represented as a single point, and the corresponding point-to-point cross-correlation with the center pixel is performed with respect to time lag $\tau$. This process is repeated for the mean and maximum horizontal travel distances, and the cross-correlations of each range are kept separate. We perform a 2x2 spatial average of the cross-correlations to reduce noise, and the resulting spatial dimensions are 64x64 with a resolution of 0.24 degrees per pixel.

The process for obtaining the directional mean travel time perturbations for each annulus group differs depending on the definition one chooses for the travel time, though both processes begin the same. Within a given range, the cross-correlations for positive time lags and a given direction are averaged with the cross-correlations for negative time lags in the opposite direction (e.g. positive time lags in the North direction and negative time lags in the South direction). The resulting cross-correlations are averaged over the x and y directions to obtain a reference cross-correlation for the given direction and annulus range.

For the Gabor wavelet derived travel time, we use a non-linear least squares algorithm to fit the reference cross-correlations to a Gabor wavelet given by
\[
\Psi(\tau) = A \cos\left(\omega(\tau-\tau_{\text{phase}})\right)\exp\left[-\dfrac{\left(\tau-\tau_{\text{group}}\right)^2}{2\gamma^{2}}\right],
\]
where $A$ is the amplitude, $\omega$ is the frequency, $\gamma$ determines the width of the Gaussian envelope, and $\tau_{\text{phase}}$ and $\tau_{\text{group}}$ are the phase and group travel times, respectively. We consider the phase travel time to represent the travel time of an acoustic wave, as it is much less susceptible to noise than the group travel time \citep{tomog}. The resulting travel times for each annulus range and direction are used to shift the cross-correlations so that the corresponding reference phase travel times are identical, and the cross-correlations are averaged over the annulus range. The resulting parameters of the initial reference fit are used for the pixel-by-pixel fit, which is performed for each direction. The reference travel time is subtracted from a given pixel's travel time to obtain the travel time perturbation, and the perturbed times for opposing directions are then averaged to obtain the mean travel time perturbations $\delta\tau_{\text{mean}}^{(i)}$ used to compute the anisotropy parameter $A$ and orientation of the subsurface horizontal magnetic field $\alpha$.

We begin the procedure for the obtaining travel times in the \cite{GB02} definition by completing the previous steps up to the pixel-by-pixel fit. We also define a reference cross-correlation, $C_{\text{ref}}$, for the entire region by averaging the directional reference cross-correlations. We then define $X_{\pm}(C,t)$, a function of the directional cross-correlation of a given pixel $C(x,y,t)$ and time lag $t$, given by
\[
X_{\pm}(C,\tau) = \int_{-\infty}^{\infty}\left[C(x,y,\tau^\prime)-C_{\text{ref}}(\tau^{\prime}\mp\tau)\right]^2f(\pm \tau^\prime)d\tau^\prime,
\]
where $f(\mp \tau)$ is a window function used to isolate the positive or negative time lags, defined to be unity in the region of $\tau_{\text{phase,ref}}\pm5$ minutes and zero otherwise. The above function is essentially the least-squared difference between a given cross-correlation and the reference cross-correlation. The travel time is then defined to be the time lag which minimizes the above function, where the backward travel time deviation $\delta\tau_{-}$ minimizes $X_{-}(C,\tau)$ and the forward travel time deviation $\delta\tau_{+}$ minimizes $X_{+}(C,\tau)$. The mean travel time perturbations $\delta\tau_{\text{mean}}^{(i)}$ are obtained by averaging the forward and backward travel time deviations, and $A$ and $\alpha$ are computed with equations \ref{aniso} and \ref{angle}, respectively.

\subsection{Numerical Simulations}

To verify the equations, we first apply our measurement scheme to the simulations developed by \cite{PKV}, where a point-source is used to generate acoustic waves propagating through a uniform, inclined magnetic field. The background model is derived from the standard solar model, as described in \cite{ssmodel}, which reproduces the observed p- and f-mode dispersion relation. Several configurations of the magnetic field are considered in \cite{PKV}, though we select the $B_{0}=$1900 G simulation where the field is inclined $30^{\circ}$ relative to normal and horizontal component oriented in the $+\hat{y}$-direction, as the amplitude and inclination are closer to the conditions expected within a sunspot's umbra. Additionally, the authors examine how the line-of-sight affects the measured travel times, though this is beyond the scope of our work. For simplicity, we therefore measure the travel times using only the vertical component of the simulated photospheric velocity.

\begin{figure}[ht!]
\begin{center}
    \includegraphics[width=\linewidth]{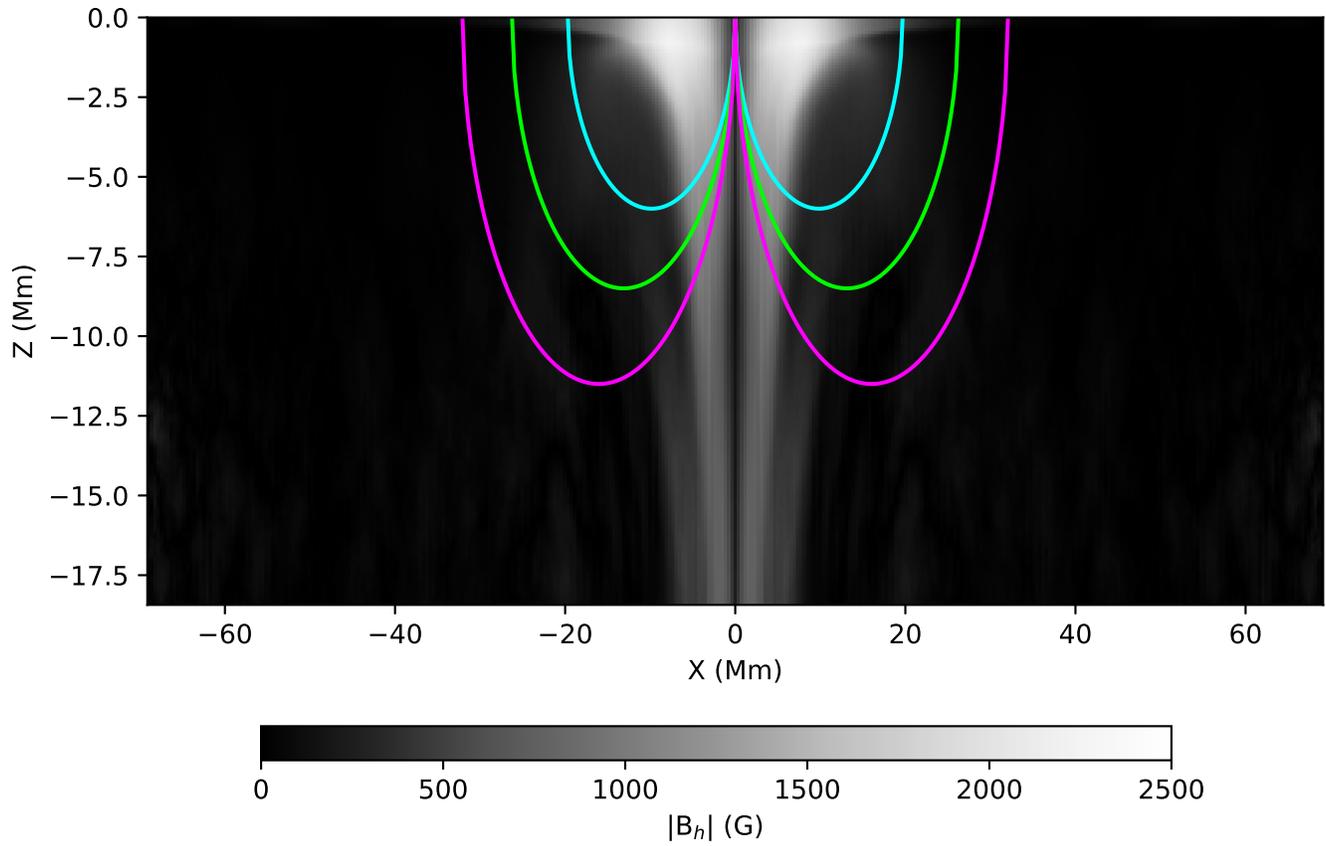}
    \caption{Path of acoustic waves in the surface-focus procedure through the sunspot simulation for measuring the anisotropy at (x,y)=0 Mm. The mean path traveled by waves belonging to annulus group 4 is highlighted in cyan, annulus group 5 in green, and annulus group 6 in magenta.}\label{rays}
\end{center}
\end{figure}

The numerical simulation of a sunspot used in this work was developed by \cite{Rempel}, where the radiative MHD equations are solved in a domain with extent 98.304 Mm in each horizontal direction and 12.432 Mm in the vertical direction. The initial condition of the simulation is an axisymmetric flux concentration of $9\times10^{21}$ Mx, which produces a $6\times10^{21}$ Mx sunspot. The data made publicly available are a time series of the vertical velocity at the model's photosphere (optical depth $\tau=0.01$ layer), the time-averaged photospheric continuum intensity and magnetic field components, as well as azimuthally- and time-averaged vertical structure of the sunspot's radial (in terms of polar coordinates) and vertical magnetic fields, among other quantities. The time series begins after 50 hours of model time, with a duration of 25 hours and cadence of 45 seconds, and the horizontal resolution of the data is 0.384 Mm per pixel. The average ray path for each annulus group is overlaid on the azimuthally averaged horizontal magnetic field magnitude in Figure \ref{rays}.

\section{Results}

\subsection{Validation of the Measurement Scheme in a Uniform, Inclined Magnetic Field}

Using the simulation developed by \cite{PKV}, we aim to validate two aspects of our measurement scheme. First, we will show that the derived equations can be used to accurately determine the orientation of the horizontal magnetic field, and by extension the travel time anisotropy which serves as a proxy for the field's magnitude. As we know the parameters of the magnetic field, we immediately know the correct orientation of the subsurface horizontal field, and can use the ray equations to determine the expected travel time anisotropy. Second, we will show that our measurements of the field's orientation and travel time anisotropy are accurate regardless of the measuring scheme's orientation. This second criterion shows that our measurements remain accurate even when the magnetic field is not well-aligned with a particular ray in the measurement scheme. We accomplish this by varying the orientation of the scheme, initially oriented such that the ray corresponding to $\delta\tau_{\text{mean}}^{(1))}$ is in the $+\hat{y}$- or NS-direction, in increments of 1$^{\circ}$ counter-clockwise.

\begin{figure}[ht!]
\begin{center}
    \includegraphics[width=\linewidth]{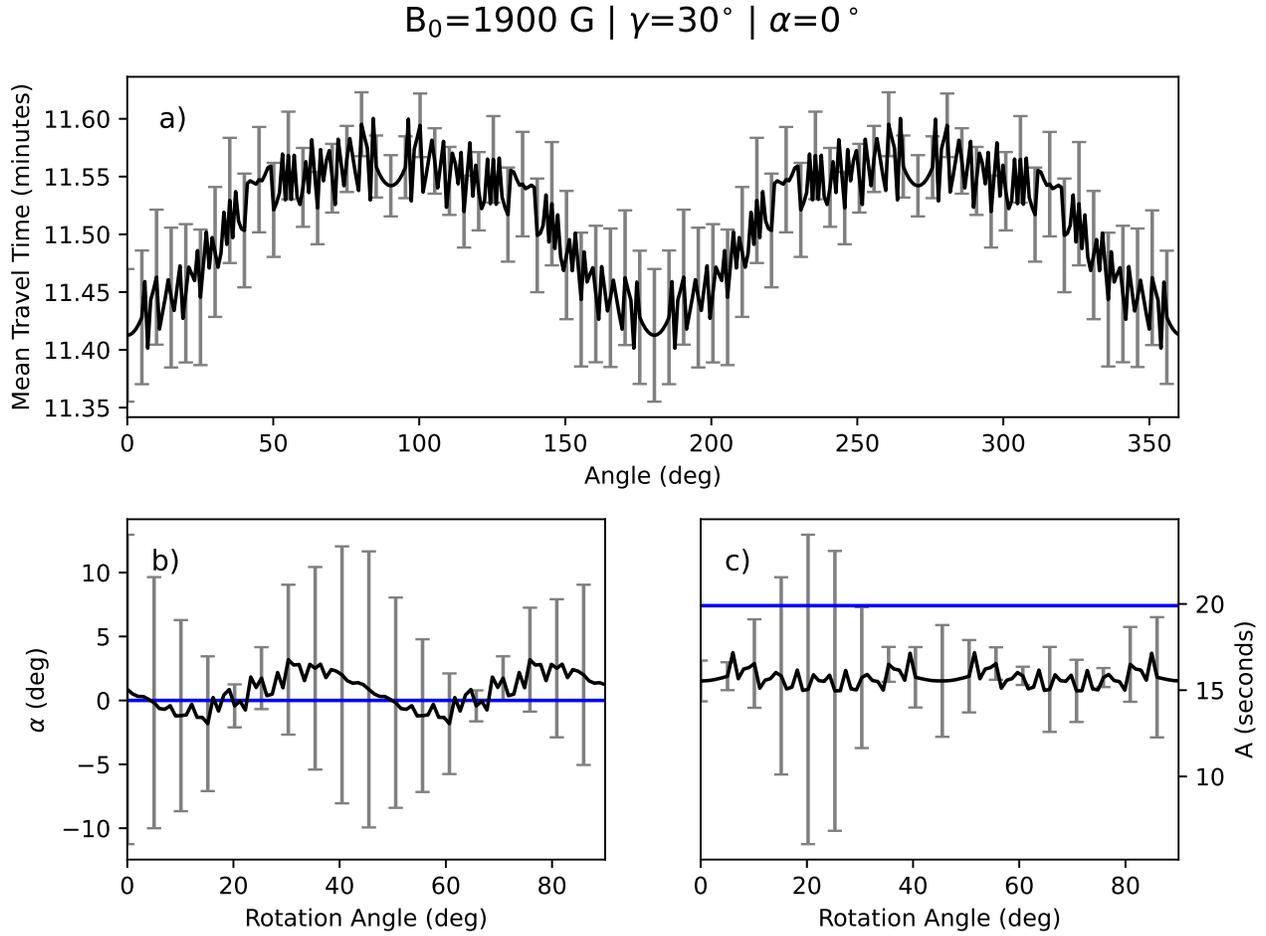}
    \caption{Validation of the measurement scheme. Panel a contains the mean phase travel times determined by the Gabor wavelet fit, panel b contains the field's inferred azimuth, and panel c contains the resulting travel time anisotropy. The blue line in panel b shows the field's true azimuth of $0^{\circ}$ relative to the $+\hat{y}$- or NS-axis, and the blue line in panel c shows the expected travel time anisotropy from solving the ray equations for the specified field's parameters.}\label{valid}
\end{center}
\end{figure}

The results of this validation are displayed in Figure \ref{valid}, with Figure \ref{valid}a showing the mean phase travel times obtained by the Gabor wavelet fit. The associated uncertainties are determined by the standard deviation of each travel time from the respective fitting. The inferred azimuth of the field and travel time anisotropy are shown in Figures \ref{valid}b and \ref{valid}c, respectively, as functions of the rotation of the measurement scheme. Figure \ref{valid}b shows that the scheme is able to accurately derive the horizontal field's azimuth to within a few degrees. Variations in the mean phase travel time are responsible for the small-scale fluctuations, and the larger-scale variations of period 45$^{\circ}$ likely originate from the features seen in the mean phase travel times with the same period. The travel time anisotropy in Figure \ref{valid}c is much more consistent than the derived azimuth, showing only the small-scale variations also present in the derived azimuth. The average anisotropy is around 15.5 seconds, close to the amplitude predicted by the ray equations of 19.9 seconds. It is not immediately clear where this discrepancy originates, though it should be noted that the computed anisotropy is highly dependent on the accuracy of numerical integration. Moreover, the uncertainties reported in Figures \ref{valid}b and \ref{valid}c are computed by considering four cases where the original mean phase travel time uncertainties are either added or subtracted from the differences in Equations \ref{aniso} and \ref{angle}. The reported uncertainties are for the "worst case" scenario, which represent the greatest uncertainty. With these factors in mind, it is reasonable to consider the derived measurement scheme to be accurate.

\subsection{Numerical Sunspot Model}

We then proceed to test the measurement technique on a more realistic simulation developed by \cite{Rempel} of a sunspot, where the magnitude of the subsurface horizontal magnetic field is known. Figure \ref{spotsim}a shows the surface magnetic field azimuth and Figure \ref{spotsim}b shows the magnitude of the surface horizontal magnetic field. It should be noted that the time-distance measurement of the subsurface azimuth cannot distinguish between polarities, so the model's azimuth has been ambiguated to make a more clear comparison. The time-distance measurements were made using annulus group 4, with turning points between 5 and 7 Mm beneath the photosphere. This choice is necessitated by the small size of the simulation's domain relative to the horizontal travel distance, which for annulus group 4 approaches 22 Mm. Deeper measurements can be made, at the cost of a severely restricted field of view. The measurements shown in Figure \ref{spotsim} are derived from the GB02 travel time definition, which is much less sensitive to noise in the cross-correlations.

The outer 10 Mm, relative to the center of the sunspot, of the measured azimuth in Figure \ref{spotsim}c corresponds fairly well with the surface azimuth, though with some additional noise closer to the inner 10 Mm. Interestingly, the inner 10 Mm shows a $90^\circ$ counter-clockwise rotation of the azimuth relative to the surface azimuth, which suggests a curl-like structure of the horizontal field in contrast with the expected radial orientation. This inner region also has surprising characteristics in the measured travel time anisotropy (Figure \ref{spotsim}d), which is severely depressed in comparison to the outer 10 Mm of measurements. Additionally, the travel time anisotropy appears not to coincide with the location of maximum horizontal magnetic field, either on the surface (Figure \ref{spotsim}b) or at depths corresponding to the turning points. The structure of the model's subsurface horizontal magnetic field remains approximately constant with depth, though the maximum magnitude decreases to about 1100 G at these layers.

\subsection{Sunspots Observed by HMI}

The sunspots studied in this work are chosen based on several criteria: the sunspot's size and shape remain relatively constant over the observation period, that the magnitude and orientation of the magnetic field also remain relatively constant, and the region surrounding the sunspot be free of significant magnetic features. The first two criteria ensure that the approximations made in Section \ref{deriv} are valid, and the third criterion aids in minimizing noise in the measurements. We then select the sunspots associated with ARs 12218, 12786, and 12794, all of which are unipolar and roughly axisymmetric. The unsigned magnetic flux reported for these sunspots in the following sections is computed with a threshold of 200 G, to reduce contributions from the surrounding magnetic features.  We examine the subsurface azimuth and travel time anisotropy for the annulus groups 4, 5, and 6, corresponding to turning points between 5-7 Mm, 7-10 Mm, and 10-13 Mm beneath the photosphere, respectively. Additionally, we compare the measurements made using the Gabor wavelet derived travel times with the GB02 derived travel times. The observations of the azimuth of the surface magnetic field and the magnitude of the horizontal and LOS components have the original 0.12 degrees per pixel resolution, and are averaged from a time series spanning the duration of the Dopplergram measurements.

\begin{figure}[ht!]
\begin{center}
    
    \includegraphics[width=\linewidth]{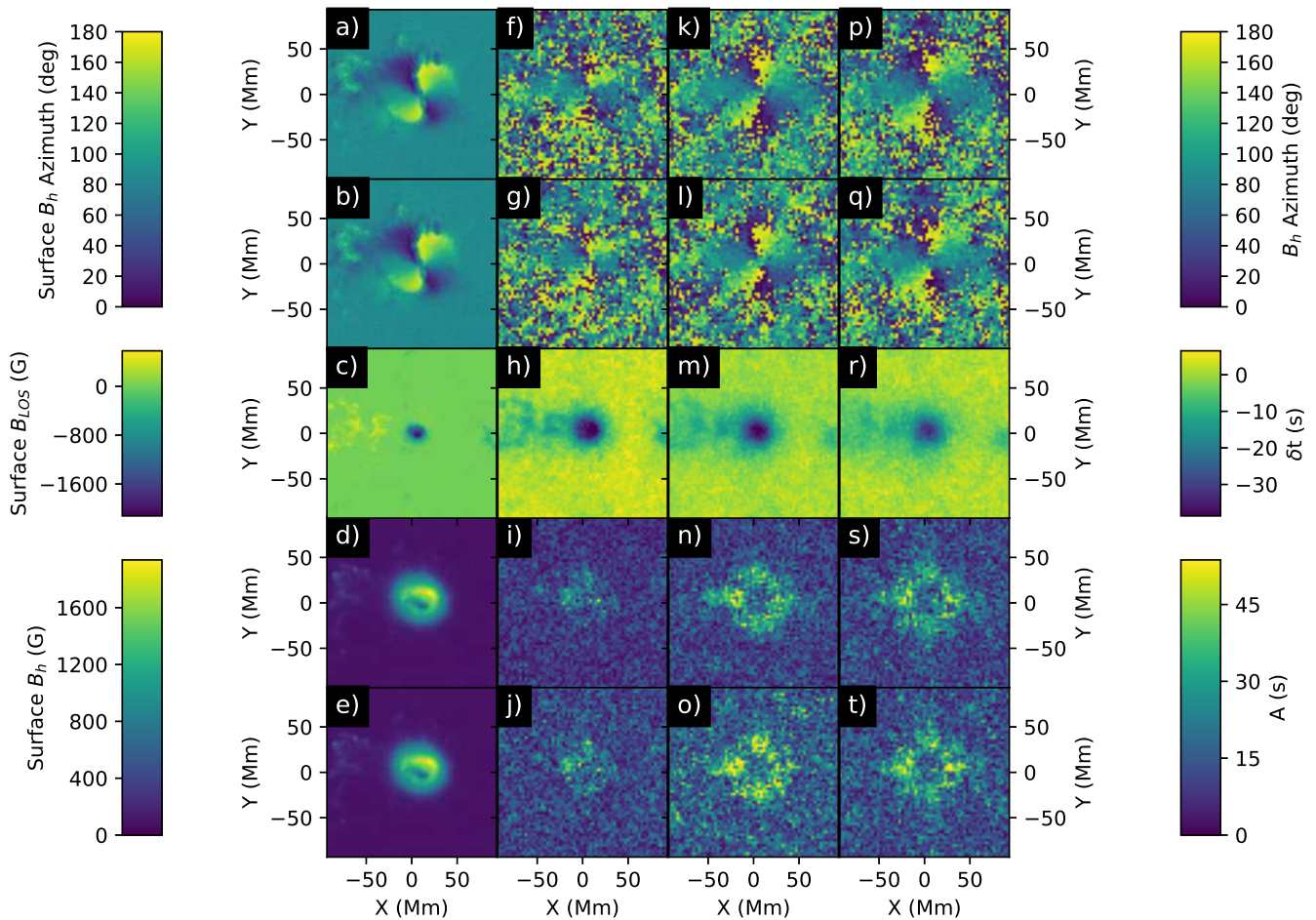}
    \caption{Comparison of AR 12218's surface magnetic field with measurements of the subsurface field. The active region's surface azimuth is shown in panels a and b, the line-of-sight magnetic field in panel c, and the magnitude of the horizontal surface magnetic field is shown in panels d and e. The measured subsurface azimuth is shown in rows 1 and 2 and travel time anisotropy in rows 4 and 5. Measurements in rows 1 and 4 are derived from the GB02 travel time method, and measurements in rows 2 and 5 are derived from the Gabor wavelet phase travel time method. Row 3 shows the mean in-out travel time difference, from which wave speed perturbations can be obtained through inversion. Column 2 corresponds to annulus group 4, column 3 to annulus group 5, and column 4 to annulus group 6.}\label{12218}
\end{center}
\end{figure}

Beginning with AR 12218 (Figure \ref{12218}) of unsigned radial magnetic flux $2.11\times 10^{22}$ Mx, we see that the subsurface measurements are much noisier than those made from the sunspot model, especially for the shallower measurements of annulus group 4 (Figures \ref{12218}f,g,i,j). The size of the detected features also changes between annulus group 4 and annulus groups 5 (Figures \ref{12218}k,l,n,o) and 6 (Figures \ref{12218}p,q,s,t), from approximately the same size as the sunspot to slightly larger. Furthermore, for annulus groups 5 and 6, the regions of greatest travel-time anisotropy do not coincide with the regions of greatest horizontal magnetic field magnitude on the surface, similar to annulus group 4 for the sunspot model. The travel-time anisotropy for annulus group 4 in this case is too noisy to make such an assessment. The measurements of the subsurface azimuth derived from the GB02 travel-time definition (Figures \ref{12218}f,k,p) and the Gabor wavelet phase travel-time (Figures \ref{12218}g,l,q) are not significantly different, and show a similar amount of noise. The advantages of the GB02 travel-time definition become more apparent in the maps of travel-time anisotropy (Figures \ref{12218}i,n,s), where there is less noise than in the Gabor wavelet derived measurements (Figures \ref{12218}h,o,t). In particular, the travel-time anisotropy in Figure \ref{12218}n is especially well-resolved, and even suggests some degree of connection to the plage on the left.

\begin{figure}[ht!]
\begin{center}
    
    \includegraphics[width=\linewidth]{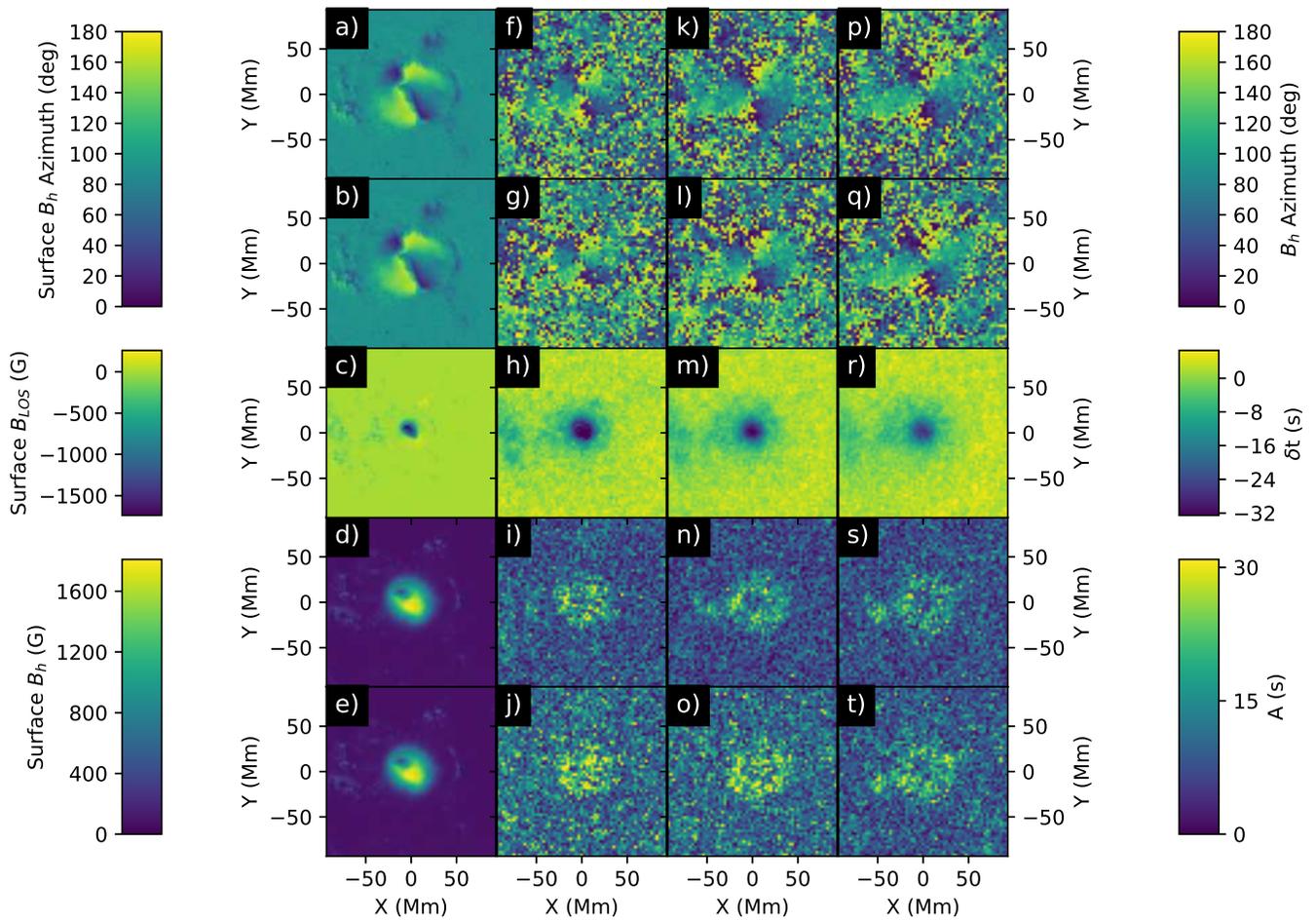}
    \caption{Same as in Figure \ref{12218}, but for AR 12786.} \label{12786}
\end{center}
\end{figure}

AR 12786 has a unique configuration, in that the magnitude of the surface horizontal magnetic field is significantly stronger on the South side than on the North, with the azimuth showing a similar amount of asymmetry. The unsigned radial flux for this region is $1.37\times 10^{22}$ Mx. In contrast to AR 12218, neither the GB02 derived (Figures \ref{12786}f,k,p) nor the Gabor wavelet derived (Figures \ref{12786}g,l,q) azimuth correspond well with the surface azimuth. In further contrast, while the maximum magnitudes of the horizontal magnetic field for AR 12218 and 12786 are roughly equal, the travel-time anisotropy for AR 12786 is nearly half of what is measured for AR 12218. In comparing their mean in-out travel-time deviations (panels h,m,r in Figures \ref{12218} and \ref{12786}), we find that the features have similar magnitude, though AR 12218's features are larger, hinting at a possible source for the discrepancy in anisotropy. Still, the anisotropy in AR 12786 is similarly well-resolved to AR 12218, and the anisotropy for annulus group 4 (Figures \ref{12786}i,j) is more pronounced than for the same annulus group in AR 12218. Again, the Gabor wavelet-derived anisotropy (Figures \ref{12786}j,o,t) displays more noise than the GB02 derived anisotropy (Figures \ref{12786}i,n,s). AR 12786 also has a plage to it's left, and the measured travel-time anisotropy in Figures \ref{12786}n, s, and t seem to indicate some connection between the plage and sunspot. The change in feature size between annulus group 4 and annulus groups 5 and 6 is not as great as for AR 12218, though the smaller size of AR 12786's sunspot may make this change less discernible.

\begin{figure}[ht!]
\begin{center}
    
    \includegraphics[width=\linewidth]{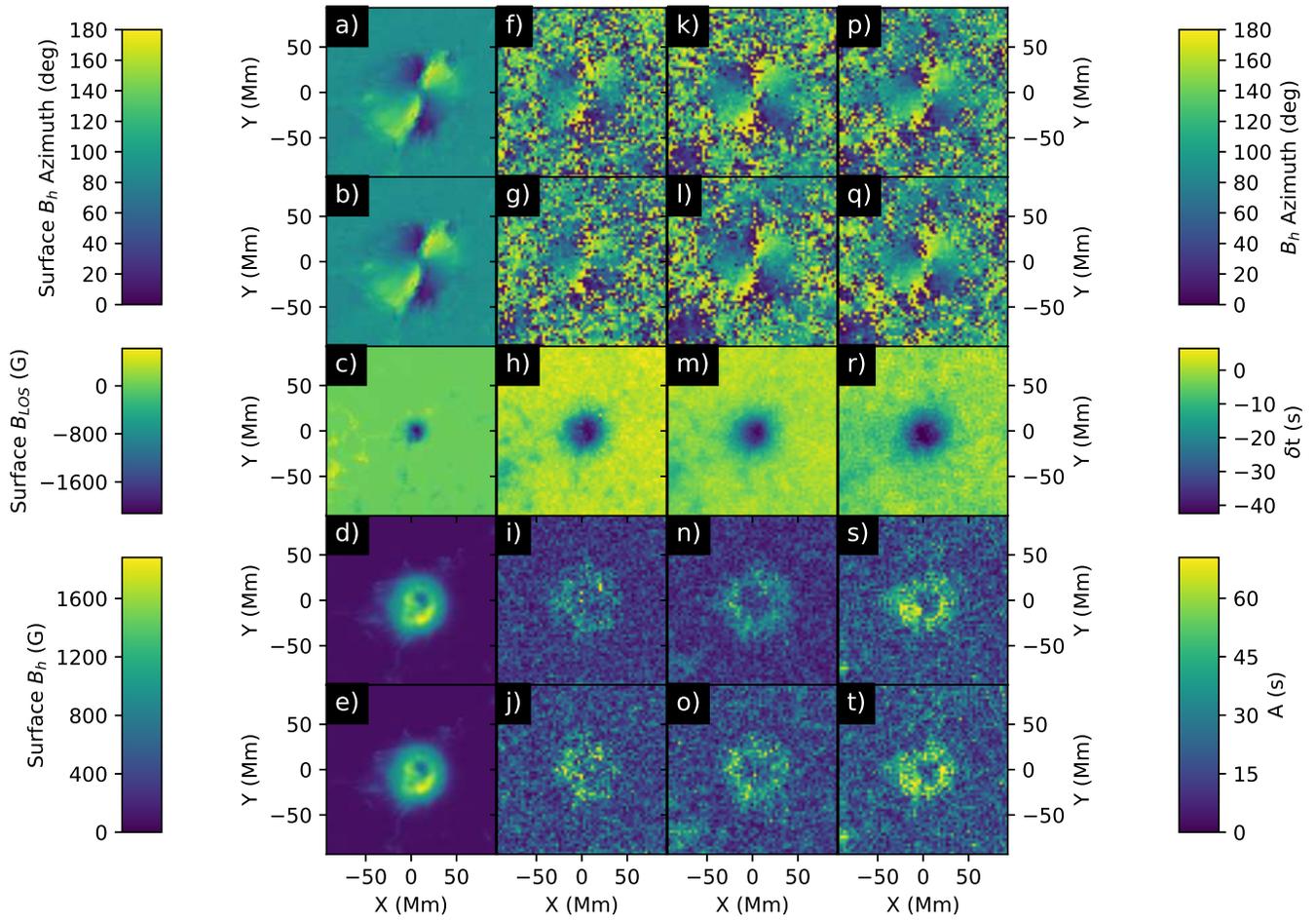}
    \caption{Same as in Figure \ref{12218}, but for AR 12794.}\label{12794}
\end{center}
\end{figure}

The final active region studied is AR 12794, having unsigned radial flux of $2.51\times 10^{22}$ Mx. This active region's surface horizontal magnetic field and azimuth are more axisymmetric than for AR 12786. Here, as with AR 12786, the features detected in the travel-time anisotropy do not change size between annulus group 4 (Figures \ref{12794}i,j) and annulus groups 5 (Figures \ref{12794}n,o) and 6 (Figures \ref{12794}s,t). In fact, the feature in annulus group 6 appears to be slightly smaller than the previous two groups. Additionally, the travel-time anisotropy derived from both the GB02 definition (Figures \ref{12794}i,n,s) and Gabor wavelet fitting (Figures \ref{12794}j,o,t) increases with depth, a phenomenon not seen with either AR 12218 or AR 12786. As with the previous two active regions, the Gabor wavelet derived travel-time anisotropy contains more noise, though annulus group 6 (Figure \ref{12794}t) has a similar noise level as the GB02 derived anisotropy for the same annulus group (Figure \ref{12794}s). The subsurface azimuth (Figure \ref{12794}f,g,k,l,p,q) is reasonably well-resolved, and annulus groups 5 and 6, in particular, indicate a unidirectional concentration of magnetic flux in the lower left corner.

\section{Discussion and Conclusion}
\begin{figure}[ht!]
\begin{center}
    
    \includegraphics[width=\linewidth]{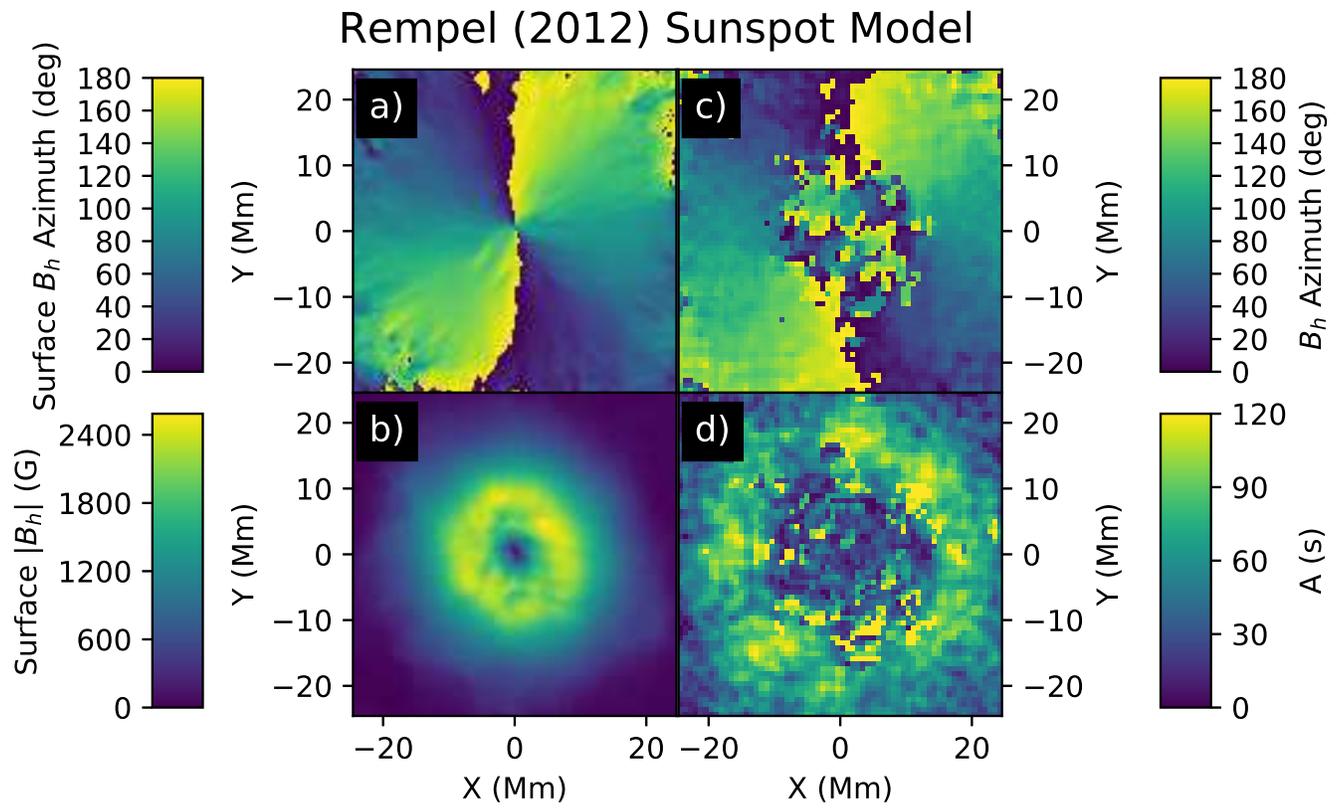}
    \caption{Comparison of the model's surface magnetic field azimuth (a) and magnitude (b) with the measured subsurface azimuth (c) and travel time anisotropy (d).}\label{spotsim}
\end{center}
\end{figure}

Perhaps the most surprising feature in the numerical simulation's measurements is the apparent $90^{\circ}$ counter-clockwise rotation of the horizontal field's azimuth (Figure \ref{spotsim}c), relative to the surface azimuth and outer 10 Mm of the measured subsurface azimuth. It is unclear if this is caused by an actual change in the subsurface azimuth, as the depth-dependent data provided only describes the magnitude of the horizontal field. This feature most likely develops as a result of the invalidity of the uniform magnetic field assumption close to the center of the sunspot, as the surface magnetic field exhibits a polarity change about the sunspot's center. The coincidence of depressed travel-time anisotropy (Figure \ref{spotsim}d) seems to support this hypothesis, where the feature is much larger than the region on the surface where the horizontal field is close to zero. In both cases, the anti-symmetric term in equation \ref{antisymm} may produce a non-zero, opposing contribution to the travel-time where the magnetic field is non-uniform.

While such a rotation of the azimuth is not observed in any of the studied active regions, the significantly coarser resolution in combination with the noise level may obfuscate this feature. Indeed, the azimuth inversion lines, where the azimuth changes from $180^{\circ}$ to $0^{\circ}$, does not extend to the center of the sunspot in any of the active region measurements, similar to the azimuth in Figure \ref{spotsim}c. However, based on the results of the numerical simulation we should expect the region avoided by the azimuth inversion lines to be of the same size as that where the travel-time anisotropy is depressed, which is only the case for ARs 12786 and 12794 in the deepest measurements of annulus group 6.

Another confounding aspect of our results is the significantly reduced magnitude of the travel-time anisotropy for AR 12786, as compared to ARs 12218 and 12794. Though the actual magnitude of the subsurface horizontal magnetic field is unknown, each of the active regions has similar surface magnitude, so we should expect their travel-time anisotropies to also have similar magnitudes. A possible contribution to the reduction in travel-time anisotropy magnitude is AR 12786's slightly smaller spatial extent as compared to the other active regions. This would also explain the greater maximum of travel-time anisotropy in AR 12794 compared with AR 12218, which is moderately smaller, as an acoustic wave traveling through AR 12794 would encounter more magnetic flux and therefore have a more perturbed travel-time.

\begin{figure}[ht!]
\begin{center}
    
    \includegraphics[width=\linewidth]{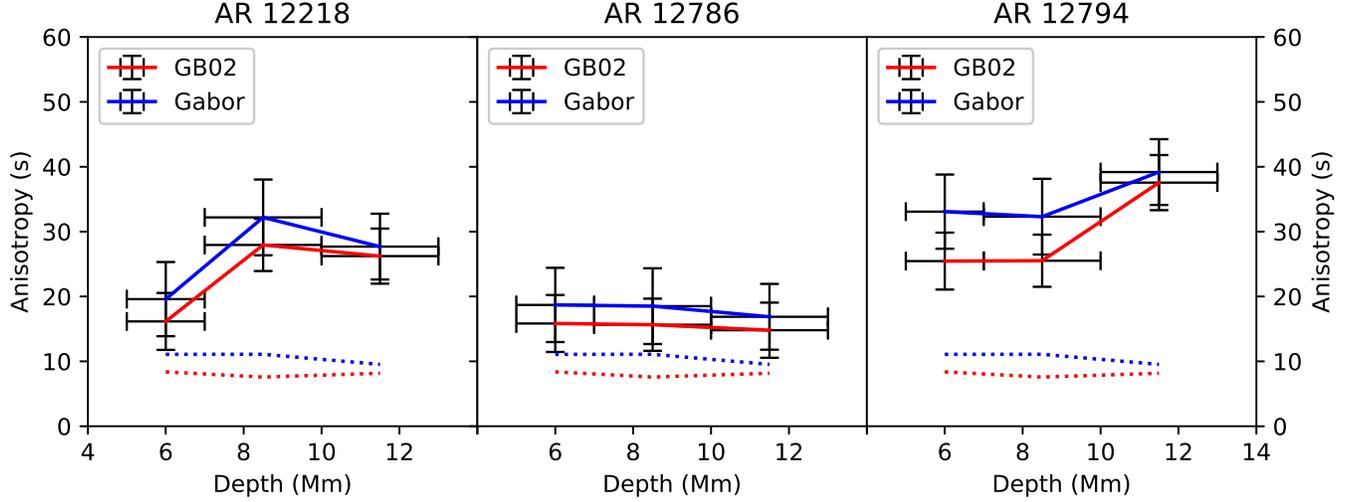}
    \caption{Spatially averaged anisotropy as a function of depth. The measurements from each annulus group are assumed to sample the middle of their respective depth ranges. The noise floor derived from quiet sun measurements is shown for each fitting method as a dotted line.}\label{depth}
\end{center}
\end{figure}

Finally, we examine how the measured anisotropy changes over depth for each of the observed active regions in Figure \ref{depth}. The anisotropy is averaged over an annulus with inner radius 10 Mm and outer radius 33 Mm, roughly corresponding to the size and shape of the detected features. We determine the noise floor (shown as dotted lines in Figure \ref{depth}) for these measurements by taking the weighted mean of travel-time anisotropy measured in a quiet Sun region, where the weights are the inverse of the chi-squared residuals for the Gabor fitting method, and the least squared error for the GB02 method. The uncertainty for each method is computed by the weighted standard deviation of the quiet Sun anisotropies, with the same weights as previously mentioned. We find that there is little, if any, similarity between depth-dependent anisotropies in the active regions. AR 12218 shows an increase in anisotropy between annulus groups 4 and 5, which levels off at annulus group 6. The anisotropy detected in AR 12786 is nearly constant with depth, with a small decrease in the transition from annulus group 5 to 6, and the anisotropy detected in AR 12794 only increases between annulus groups 5 and 6. Clearly, the radial structure of the anisotropy is highly dependent on local conditions. 

In comparing the results derived from the GB02 travel-time definition with those derived from the Gabor wavelet fitting, we find that the GB02 derived results outperform in almost every instance. We expect this since the GB02 method is far less sensitive to noise in the cross-correlations, in part because GB02 travel-time depends only on time lag, whereas the Gabor wavelet method requires the frequency, width of Gaussian envelope, amplitude, and group travel-time to also be included in the fitting. This may be useful in analytically reconstructing the cross-correlation function, but comes at the cost of reduced phase travel-time accuracy as the combined error of all the variables is minimized, as opposed to solely the phase travel-time as in the GB02 method.

In conclusion, we find that the results of the numerical sunspot simulation confirm that the travel-time anisotropy defined in equation \ref{aniso} is a reasonable proxy for the subsurface horizontal magnetic field, and the combination of travel-times in equation \ref{angle} reproduces the subsurface azimuth quite well. While the structure of the detected features does not spatially coincide with the structure of the subsurface horizontal magnetic field, this discrepancy may be explained by other terms in our derivation where the uniform field approximation fails. Despite the increased noise in the measurements, we show that detection of travel-time anisotropy in active regions is possible and consistent with what is observed on the surface. We also find that the size of a magnetized region influences the travel-time anisotropy more than expected. In future work, we will develop an inversion method to reconstruct the horizontal magnetic field from the travel-time anisotropy to provide a new, important diagnostic in the study of active regions.

\begin{acknowledgements}
We thank M. Rempel and K. Parchevsky for their publically available simulations. This work was supported by NASA grants NNX14AB70G, 80NSSC19K0268, 80NSSC19K0630, and 80NSSC20K1870.
\end{acknowledgements}

\newpage

\newpage
\newpage
\newpage
\newpage
\newpage

\newpage

\newpage
\newpage

\appendix
\section{Derivation of the Wavenmumber Perturbation Produced by Acoustic to Fast-Mode Conversion}\label{mhd_deriv}

We consider linear perturbations in the adiabatic condition (equation \ref{adi}), conservation of momentum (equation \ref{mom}) and mass (equation \ref{con}), and the induction equation (equation \ref{ind}), for a background magnetic field $\mathbf{B_{0}}$ oriented in the $\mathbf{\hat{z}}$ direction and an initially stationary plasma. Denoting the perturbed variables with $^{\prime}$, we have
\begin{equation}\label{adi}
p^{\prime}=c_{s}^{2}\rho^{\prime},
\end{equation}
\begin{equation}\label{mom}
\rho_{0}\dfrac{\partial \mathbf{v}}{\partial t} = -\nabla\left(c_{s}^{2} \rho^{\prime}\right) + \dfrac{1}{4\pi}\left(\nabla\times \mathbf{B^{\prime}}\right)\times \mathbf{B_{0}},
\end{equation}
\begin{equation}\label{ind}
\dfrac{\partial \mathbf{B^{\prime}}}{\partial t} = \nabla\times\left(\mathbf{v}\times\mathbf{B_{0}}\right),
\end{equation}
and
\begin{equation}\label{con}
\dfrac{\partial \rho^{\prime}}{\partial t} = -\rho_{0}\nabla\cdot\mathbf{v}.
\end{equation}

Considering plane wave solutions, and orienting the coordinate system so that the direction of propagation is along the $\hat{z}$ axis, i.e.
\[
\dfrac{\partial}{\partial x} = \dfrac{\partial}{\partial y} = 0 ; \dfrac{\partial}{\partial z}=ik ; \dfrac{\partial}{\partial t} = -i\omega,
\]
the following equations can be obtained from equation \ref{mom}:
\[
\left\lbrace
\begin{array}{r}
	\omega\rho_{0}v_{x} + \dfrac{B_{0}}{4\pi}kB_{x}\cos\alpha=0 \\
	\omega\rho_{0}v_{y} + \dfrac{B_{0}}{4\pi}kB_{y}\cos\alpha=0 \\
	\omega\rho_{0}v_{z}-kc_{s}^{2}\rho^{\prime} - \dfrac{B_{0}}{4\pi}kB_{y}\sin\alpha=0
\end{array}\right. ,
\]
where $\alpha$ is the angle between the magnetic field and the $\hat{z}$ direction. Expanding equation \ref{ind} yields
\[
\left\lbrace
\begin{array}{l}
	\omega B_{x}+B_{0}kv_{x}\cos\alpha = 0 \\
	\omega B_{y} + B_{0}kv_{y}\cos\alpha-B_{0}kv_{z}\sin\alpha = 0 \\
	\omega B_{z} = 0
\end{array}
\right. ,
\]
and expanding equation \ref{con} yields
\[
\omega \rho^{\prime} - k\rho_{0}v_{z} = 0 .
\]

Solving for $v_{x}$ in the $x$ component of the momentum equation, and substituting into the $x$ component of the induction equation, gives the dispersion relation for Alfv\'en waves:
\begin{equation}\label{alf}
\omega^{2}\rho_{0}-k^{2}\dfrac{B_{0}^{2}}{4\pi}\cos^{2}\alpha=0 \rightarrow \omega=kc_{A}\cos\alpha
\end{equation}
where $c_{A}=B_{0}/\sqrt{4\pi\rho_{0}}$ is the Alfv\'en speed. The Alfv\'en waves have the magnetic field as the sole restoring force.

The remaining four equations can be solved as a matrix equation to obtain the dispersion relation for magnetoacoustic waves

\[
\left(\dfrac{\omega}{k}\right)^2 = \dfrac{1}{2}\left(c_{A}^{2}+c_{s}^{2}\right) \pm \dfrac{1}{2}\sqrt{c_{A}^{4}+c_{s}^{4}-2c_{A}^{2}c_{s}^{2}\cos 2\alpha}.
\]
This dispersion relation has two modes: the fast mode $(+)$ and the slow mode $(-)$. Both magnetoacoustic waves have pressure and the magnetic field as their restoring forces, though these contributions add for the fast mode, and subtract for the slow mode. The above expression can be simplified by the cosine double angle identity to
\[
\left(\dfrac{\omega}{k}\right)^{2}=\dfrac{1}{2}\left[c_{A}^2+c_{s}^2\pm\sqrt{c_{A}^{4}+c_{s}^{4}-2c_{s}^{2}c_{A}^2(2\cos^{2}\alpha-1)}\right]
\]
\[
\left(\dfrac{\omega}{k}\right)^{2}=c_{\text{MHD}}^{2}=\dfrac{1}{2}\left[c_{A}^2+c_{s}^2\pm\sqrt{\left(c_{A}^{2}+c_{s}^{2}\right)^{2}-4c_{s}^{2}\left(\dfrac{\mathbf{k}\cdot\mathbf{c_{A}}}{|\mathbf{k}|}\right)^2}\right],
\]
where we have expressed the cosine term in the form of a dot product to show the connection to the magnetic term in equation \ref{dtfull}.

Re-expressing the above equation in terms of some variable $x=c_A^2/c_s^2$,
\[
c_{\text{MHD}}^{2}(x)=\dfrac{c_s^2}{2}\left[x+1\pm\sqrt{x^2+1+2x+4xk^2 \cos^{2}\alpha}\right].
\]
For a small magnetic field $\mathbf{B_{0}}$, we expect the Alfv\'en speed to be smaller than the local sound speed, so we can consider small perturbations about $x=0$. Applying a first order Taylor expansion to fast MHD speed
\[
c_{f,\text{MHD}}^{2}(x) \approx c_{f,\text{MHD}}^{2}(x=0) + x\left.\dfrac{d}{dx}\left(c_{sf,\text{MHD}}^{2}\right)\right|_{x=0}
\]
\[
c_{f,MHD}^{2}\approx c_s^{2}+\left(c_{A}^{2}-\left(\dfrac{\mathbf{k}\cdot\mathbf{c_{A}}}{|\mathbf{k}|}\right)^2\right)
\]

Given that the perturbation to the travel time is given by the following:

\[
\delta \tau=\dfrac{1}{\omega}\int_{\Gamma}\delta k ds
\]
We must now derive the perturbation to the wavenumber. We will again Taylor expand the fast MHD speed to first order, though this time using the relation $c_{f,MHD}^{2}=\omega^2/k^2$ and expand about the unperturbed wavenumber $k_0$:

\begin{align*}
c_{f,MHD}^{2}(k) &= c_{f,MHD}^{2}(k_{0})+\left(k-k_{0}\right)\left.\dfrac{d}{dk}\left(c_{f,MHD}^{2}\right)\right|_{k=k_0} \\
 & =c_{f,MHD}^{2}(k_{0})+\delta k\left.\dfrac{d}{dk}\left(c_{f,MHD}^{2}\right)\right|_{k=k_0} \\
 & = \dfrac{\omega^{2}}{k_{0}^{2}}-2\delta k \dfrac{\omega^2}{k_{0}^{3}} \\
 & = c_s^{2}-2\delta k \dfrac{\omega^{2}}{k_{0}^{3}}
\end{align*}
Where the last equality is derived from the unperturbed dispersion relation (no magnetic field). Substituting the first Taylor expansion in the above for $c_{f,MHD}^{2}$:
\[
c_{s}^{2}+\left(c_{A}^{2}-\left(\dfrac{\mathbf{k}\cdot\mathbf{c_{A}}}{|\mathbf{k}|}\right)^2\right)=c_{s}^{2}-2\delta k \dfrac{\omega^{2}}{k_{0}^{3}}
\]
\[
\delta k = -\dfrac{1}{2}\dfrac{k_{0}^{3}}{\omega^{2}}\left(c_{A}^{2}-\left(\dfrac{\mathbf{k}\cdot\mathbf{c_{A}}}{|\mathbf{k}|}\right)^2\right)
\]

Substituting the above into the expression for the travel time perturbation:

\begin{align*}
\delta\tau &=\dfrac{1}{\omega}\int_{\Gamma}\delta k ds \\
& = -\int_{\Gamma}\dfrac{1}{2}\dfrac{k_{0}^{3}}{\omega^{2}}\left(c_{A}^{2}-\left(\dfrac{\mathbf{k}\cdot\mathbf{c_{A}}}{|\mathbf{k}|}\right)^2\right)ds \\
& = -\int_{\Gamma}\dfrac{1}{c_{s}^{2}}\left(c_{A}^{2}-\left(\dfrac{\mathbf{k}\cdot\mathbf{c_{A}}}{|\mathbf{k}|}\right)^2\right)S ds
\end{align*}
\[
\delta\tau = -\int_{\Gamma}\dfrac{1}{2}\left[\dfrac{c_{A}^{2}}{c_{s}^{2}}-\left(\dfrac{\mathbf{k}\cdot\mathbf{c_{A}}}{|\mathbf{k}|c_{s}}\right)^2\right]S ds
\]

\bibliography{main}

\end{document}